\documentclass[prl,showpacs,byrevtex,floatfix,superscriptaddress,twocolumn]{revtex4}
\usepackage{amssymb,graphicx,afterpage}

\begin{document}

\title{\boldmath Separation of orbital contributions to the optical conductivity of BaVS$_3$\unboldmath}
\author{I. K\'ezsm\'arki}
\affiliation{Electron Transport Research Group of the Hungarian
Academy of Science and Department of Physics, Budapest University of
Technology and Economics, 1111 Budapest, Hungary}
\author{G. Mih\'aly}
\affiliation{Electron Transport Research Group of the Hungarian
Academy of Science and Department of Physics, Budapest University of
Technology and Economics, 1111 Budapest, Hungary}

\author{R. Ga\'al}
\affiliation{Ecole Politechnique Federale, Lausanne, CH-1015
Switzerland}
\author{N. Bari\^si\'c}
\affiliation{Ecole Politechnique Federale, Lausanne, CH-1015
Switzerland}
\author{A. Akrap}
\affiliation{Ecole Politechnique Federale, Lausanne, CH-1015
Switzerland}
\author{H. Berger}
\affiliation{Ecole Politechnique Federale, Lausanne, CH-1015
Switzerland}
\author{L. Forr\'o} \affiliation{Ecole Politechnique
Federale, Lausanne, CH-1015 Switzerland}

\author{C.C. Homes}
\affiliation{Condensed Matter Physics and Materials Sciences
Department, Brookhaven National Laboratory, Upton, NY 11973 USA}

\author{L. Mih\'aly}
\affiliation{Electron Transport Research Group of the Hungarian
Academy of Science and Department of Physics, Budapest University of
Technology and Economics, 1111 Budapest, Hungary}
\affiliation{Department of Physics and Astronomy, Stony Brook
University, Stony Brook, NY 11794-3800}

\begin{abstract}
The correlation-driven metal-insulator transition (MIT) of BaVS$_3$
was studied by polarized infrared spectroscopy. In the metallic
state two types of electrons coexist at the Fermi energy: The quasi
$1D$ metallic transport of $A_{1g}$ electrons is superimposed on the
isotropic hopping conduction of localized $E_g$ electrons. The
"bad-metal" character and the weak anisotropy are the consequences
of the large effective mass $m_{eff}\approx7m_e$ and scattering rate
$\Gamma\geq160$\,meV of the quasi-particles in the $A_{1g}$ band.
There is a pseudo-gap above $T_{MI}=69$\,K, and in the insulating
phase the gap follows the BCS-like temperature dependence of the
structural order parameter with $\Delta_{ch}\approx42$\,meV in the
ground state. The MIT is described in terms of a weakly coupled
two-band model.
\end{abstract}
\date{\today}

\maketitle

In metals with low-dimensional electronic structure, Fermi surface
instabilities can drive the system to various kinds of
symmetry-breaking insulating ground state like a charge density wave
(CDW) or a spin density wave (SDW). Although the primary force of
these transitions is the Fermi surface instability, the ordered
state often involves a lattice distortion. According to recent X-ray
studies \cite{Fagot,Inami}, the MIT at $T_{MI}=69$\,K in the
vanadium chain compound BaVS$_3$ is also accompanied by the lowering
of the crystal symmetry with four V ions along the chains in the
doubled unit cell. However, BaVS$_3$ is far from being a
run-of-the-mill CDW or SDW system. The high-temperature phase is
categorized as a "bad metal" where the mean free path is in the
order of the V-V distance and the susceptibility follows a
Curie-Weiss law. Below the MIT, the susceptibility and the
electronic entropy drop with no signs of long range magnetic order
down to a subsequent transition at $T_x\approx30$\,K
\cite{Nakamura2000}.

According to ARPES experiments \cite{Mitrovic} and band structure
calculations \cite{Mattheiss,Whangbo2002,Whangbo2003,Lechermann} the
wide-band $1D$ electron system coexists with quasi-localized $3D$
electrons around the Fermi energy. The spin degree of freedom of the
localized electrons, the coupling between the localized and
delocalized electrons, and the possibility of orbital ordering make
the system more complex. Although the suppression of the MIT line
and an adjacent quantum-critical region has been mapped by
high-pressure experiments \cite{Booth,Kezsmarki2001,Forro2000},
there is still no generally accepted explanation for the MIT in this
compound. The goal of the present optical study was to explore the
anisotropic electronic excitations above and below the MIT.

Above $T_{MI}=69$\,K the unit cell contains two formula units and
the formal valence of the vanadium ion is $3d^1$. The two electrons
per unit cell are distributed between two broad bands of $A_{1g}$
symmetry, derived from vanadium $d_{z^2}$ orbitals with strong
overlap along the chains ($c$ axis), and four narrow bands of $E_g$
symmetry, originating mostly from $e(t_{2g})$ orbitals
\cite{Mattheiss,Whangbo2002,Whangbo2003,Lechermann}.  (For the
structure of the material and labeling of the electronic orbitals
see \textit{e.g.} Ref. \onlinecite{Mihaly}.) The electronic band
structure in local density approximation shows that the two $A_{1g}$
subbands (the "blue" bands in Fig. 1 of
Ref.~\onlinecite{Lechermann}) are roughly symmetric to the Fermi
energy with bonding- and antibonding-like character, hereafter
referred to as $A_{1g}$ and $A_{1g}^*$, respectively. They have
large dispersion along the $\Gamma Z$ line of the Brillouin zone and
the low-lying $A_{1g}$ branch crosses the Fermi level at the $Z$
point implying strongly anisotropic conduction with the $c$ axis as
the best conducting direction. Recent ARPES experiments also
indicate that two types of V bands ($A_{1g}$ and $E_g$) cross the
Fermi energy \cite{Mitrovic}. Large on-site Coulomb repulsion is
expected to introduce a correlation between electrons and lead to an
approximately half-filled $A_{1g}$ band \cite{Lechermann}.

Polarized reflectivity was studied on a high-quality oriented mosaic
sample of BaVS$_3$ in a photon energy range of
$\hbar\omega=3$\,meV$-$$3.2$\,eV. The optical conductivity obtained
by Kramers-Kronig transformation is denoted by $\sigma_{\parallel}$
and $\sigma_{\perp}$ for polarizations parallel and perpendicular to
the chain direction, respectively. The dc resistivity
$\rho_{\parallel}$ was also measured on a high-quality BaVS$_3$
crystal in an extended temperature range up to $T=650$\,K. The
$\rho_{\parallel}(T)$ curve is plotted in Fig.~1 together with the
points derived from the optical conductivity according to
$1/\sigma_{\parallel}(\omega\rightarrow0)$. The fairly good
agreement between the two quantities confirms the validity of the
data processing.

\begin{figure}[h!]
\includegraphics[width=2.8in]{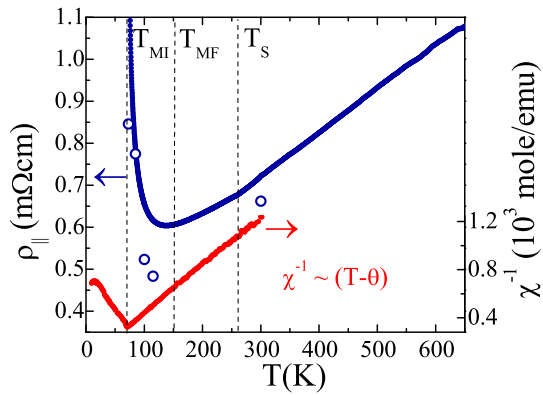}
\caption{(Color online) Temperature dependence of the resistivity
$\rho_{\parallel}$ (blue) and the inverse magnetic susceptibility
$\chi^{-1}$ (red). Open circles corresponds to the zero-frequency
extrapolation of the optical spectra according to
$\rho_{\parallel}\equiv1/\sigma_{\parallel}(\omega\rightarrow0)$.}
\end{figure}

Fig.~2 shows the reflectivity and the wide-energy range conductivity
spectra at representative temperatures for the two polarization
directions. Above $\sim1$\,eV the optical conductivity is dominated
by interband transitions leading to a broad, featureless
contribution to both $\sigma_{\perp}(\omega)$ and
$\sigma_{\parallel}(\omega)$. This continuum of absorption is
approximately isotropic and assigned to various transitions between
strongly dispersive bands, including S$(3p)\rightarrow$V$(3d)$
electron transfer, with no particular preferred direction for the
dipole matrix element. On this featureless background two peaks
appear in $\sigma_{\parallel}(\omega)$ around $1.25$\,eV and
$2.5$\,eV. Based on the band structure calculations
\cite{Mattheiss,Whangbo2002,Whangbo2003,Lechermann} we assign the
first peak to $E_g\rightarrow A_{1g}^*$ and $S(\pi_z^*)\rightarrow
E_g$ transitions. The former involves electron transfer between
neighboring V ions and has large dipole matrix elements only for
$E\parallel c$ since the $V-V$ distance is $3$ times smaller along
the chains than in the $ab$ plane. The joint density of states is
also large, due to the flat dispersion of the bands between the
$\Gamma - C - Y$ points. We note here that $S(\pi_z^*)\rightarrow
A_{1g}^*$ process may also give rise to similar anisotropic
contribution probably corresponding to the smaller peak at
$\sim2.5$\,eV \cite{Whangbo2002}.

\begin{figure}[h!]
\includegraphics[width=2.8in]{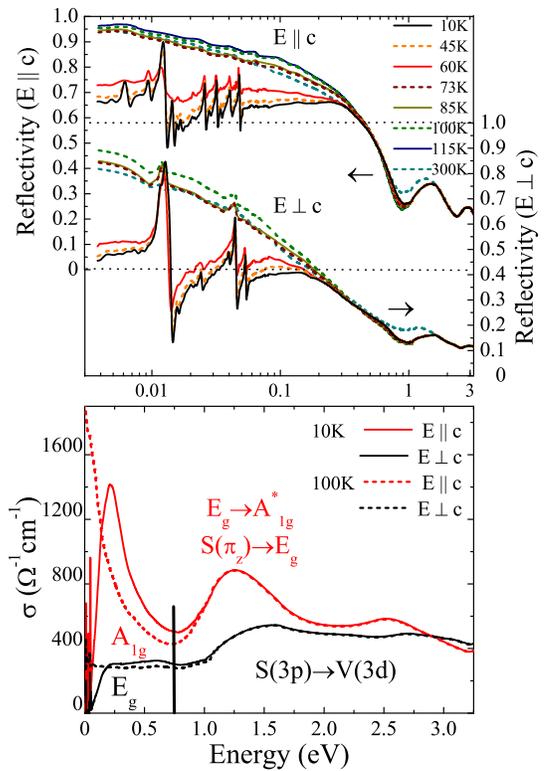}
\caption{(Color) Upper panel: Reflectivity spectra of BaVS$_3$ at
temperatures below $300$\,K for $E\parallel c$ and $E\perp c$. Lower
panel: Optical conductivity at $T=10$ and $100$\,K with polarization
parallel (red) and perpendicular (black) to the $c$ axis. Labels
show the assignment of the different contributions.}
\end{figure}

The analysis of the gap profile and the temperature dependence of
the charge gap has been widely used to categorize the
correlation-driven MITs \cite{imada}. In the low-temperature phase
of BaVS$_3$, there is a well-defined optical gap for both
$E\parallel c$ and $E\perp c$. Gap values were determined by
linearly extrapolating the steeply increasing edge of the spectra
and taking the zero conductivity intercept as indicated in the
enlarged plot of the low-energy region in Fig.~3. Below $50$\,meV
strong phonon lines are visible: Several modes are exclusive to only
one of the two polarizations, indicating the excellent orientation
of the sample. Note that apart from the phonons, the optical
conductivity vanishes below the gap energy.

The charge gap in the ground state is
$\Delta_{ch}=42\pm5$\,meV$=470\pm 60$\,K (in our convention
$\Delta_{ch}=2\Delta=E_g$). This yields an anomalously large gap
ratio, $\Delta_{ch}/T_{MI}\approx6$. Nevertheless, the temperature
dependence of the gap is close to a BCS-like behavior and matches
well that of the structural order parameter (amplitude of the
tetramerization along the V chains) as illustrated in Fig.~4(a).
Gaps for the two polarizations are close, though the values
determined from $\sigma_{\perp}(\omega)$ are somewhat higher.
Recently, Mitrovic et al. \cite{Mitrovic} estimated the ground-state
gap by the leading-edge shift of the high-resolution ($15$\,meV)
photoemission spectra and found $\Delta_{ch}=60-70$\,meV. The
estimates based on resistivity measurements
\cite{Mihaly,Graf,Nakamura1994} fall in a similar range, but they
show large variation likely due to the different quality of the
samples. Our optical study allows the first direct and precise
observation of the charge gap in BaVS$_3$.

\begin{figure}[h!]
\includegraphics[width=2.8in]{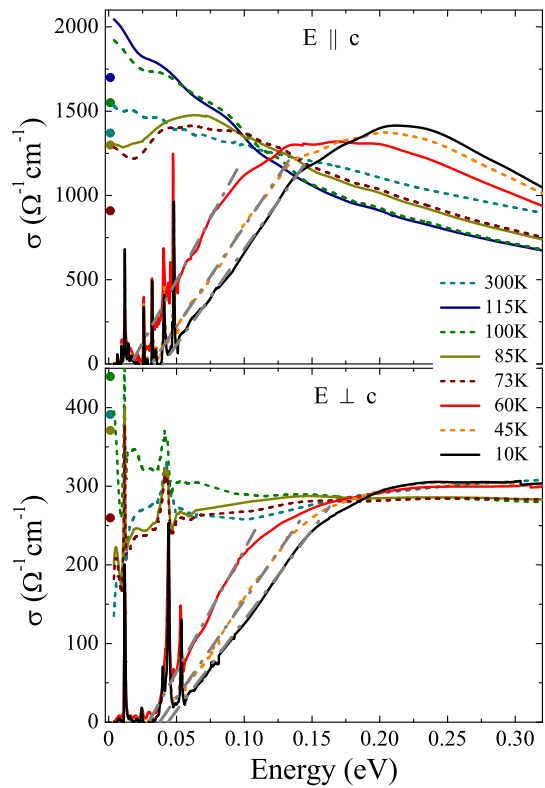}
\caption{(Color) Low-energy optical conductivity for $E\parallel c$
(top) and $E\perp c$ (bottom). Full circles represent the dc
conductivity at the same temperatures obtained from resistivity
measurements. For $T<T_{MI}$ the extrapolation to estimate the gap
energy is shown by dash dotted grey lines. Note that the scales for
$\sigma_{\parallel}$ and $\sigma_{\perp}$ differ by a factor of
$\sim5$.}
\end{figure}

Above the MIT a finite low-energy conductivity develops. For
polarization along the chains ($E\parallel c$) the phonon lines
disappear, as expected due to the screening by conduction electrons.
However, a pseudo-gap feature (a reduced conductivity at low
frequencies) can be still discerned at $T=73$\,K and $85$\,K and the
onset of coherent conduction is observed only at higher
temperatures. The presence of the pseudo-gap is in agreement with
several observations related to the $1D$ fluctuations in this
material, including the diffuse X-ray scattering\cite{Fagot} seen
below $\sim160$\,K, the upturn both in the resistivity (see Fig.~1)
and the Hall coefficient\cite{Booth}, and the appearance of an extra
contribution in the thermoelectric power\cite{Forro} below
$\sim120-140$\,K. According to the well-established quasi $1D$
scenario, the MIT is suppressed by fluctuations, and the "mean
field" transition temperature is much higher than $T_{MI}=69$\,K,
probably in the $T_{MF}=120-160$K range. Indeed, with $T_{MF}$ the
gap ratio is close to the BCS value of
$\Delta_{ch}/T_{MF}\approx3.5$. In quasi $1D$ compounds exhibiting
charge-ordering Peierls transition (Ta(Se$_4$)$_2$I,
K$_{0.3}$MnO$_3$ \cite{degiorgi} and NbSe$_3$ \cite{perucchi}),
opening of the charge gap is also observed in the ordered phase.
However, unlike in BaVS$_3$, in these materials there is always a
residual conductivity below the gap energy, attributed to $1D$
fluctuations.

The metallic conduction with a Drude-like energy dependence is
recovered along the chains around $100$\,K, where
$\rho_{\parallel}(T)$ has a minimum (see Fig.~1). At elevated
temperatures the coherence is gradually lost and at room temperature
the separation of a coherent contribution becomes ambiguous. In the
metallic phase the low-frequency optical conductivity perpendicular
to the $c$ axis is $\sim5$ times smaller than parallel to it, in
agreement with the dc anisotropy
$\sigma_{\parallel}/\sigma_{\perp}\approx4$ \cite{Mihaly,Mattheiss}.
Furthermore, $\sigma_{\perp}(\omega)$ is practically independent of
the frequency, apart from the phonon resonances that are still
visible at room temperature. The low conductivity, the absence of
coherence peak and the ineffective screening of phonon modes all
imply an incoherent hopping conduction in the $ab$ plane.

We interpret the low-frequency optical data in terms of a "two-band"
model. The $A_{1g}$ band has strong dispersion along the $c$
direction, and carries the electrons responsible for the metallic
character. The conductivity due to these carriers is expected to be
strongly anisotropic - we will assume that their contribution is
negligible in the directions perpendicular to $c$. On the other
hand, the $E_g$ electrons are localized and they form flat bands.
Their hopping conduction is essentially isotropic, since the
molecular orbitals responsible for these bands are tilted relative
to the principal crystallographic axes (their lobes point to the
face and the edge center of the sulfur octahedra), and the orbital
overlaps do not prefer any particular direction. Based on these
simple assumptions, one can deduce the conductivity of the metallic
band as $\sigma_{A_{1g}}=\sigma_{\parallel}-\sigma_{E_g}$ where
$\sigma_{E_g}=\sigma_{\perp}$. The resulting
$\sigma_{A_{1g}}(\omega)$ can be well fitted by a single Drude term,
as presented in Fig.~4(b) for $T=115$\,K. The effective number of
electrons contributing to the Drude peak is $N^{eff}\approx0.072$
and their scattering rate is $\Gamma\approx160$\,meV.

This interpretation is in agreement with other measurements, and
sheds a new light to the long standing puzzle of BaVS$_3$, how the
low conduction anisotropy is realized in a metal with quasi $1D$
crystal structure. The small $N^{eff}$ for the conduction band leads
to a large effective mass: Assuming $1/2$ electrons per vanadium in
the $A_{1g}$ subband\cite{Lechermann} we obtain
$m_{eff}\approx7m_{e}$. The large overall bandwidth of $A_{1g}$
electrons, estimated by band structure calculations
\cite{Mattheiss,Whangbo2002,Whangbo2003,Lechermann} and determined
by angle-resolved photoemission spectroscopy \cite{Mitrovic}, would
suggest a smaller mass. However, hybridization with S$(3p)$ subbands
renormalizes the band mass that is further increased by correlation
effects among which on-site Coulomb interaction is probably the most
relevant. Although the $A_{1g}$ electrons are metallic, their large
mass combined with short relaxation time ($\tau\approx$\,0.026ps)
yields a low conductivity characterized by a mean free path close to
the lattice constant\cite{Mihaly}. The relatively weak anisotropy
follows from the fact that the isotropic hopping conduction of the
$E_{g}$ electrons is not negligible relative to the quasi $1D$
metallic conduction of the $A_{1g}$ band.

\begin{figure}[h!]
\includegraphics[width=2.8in]{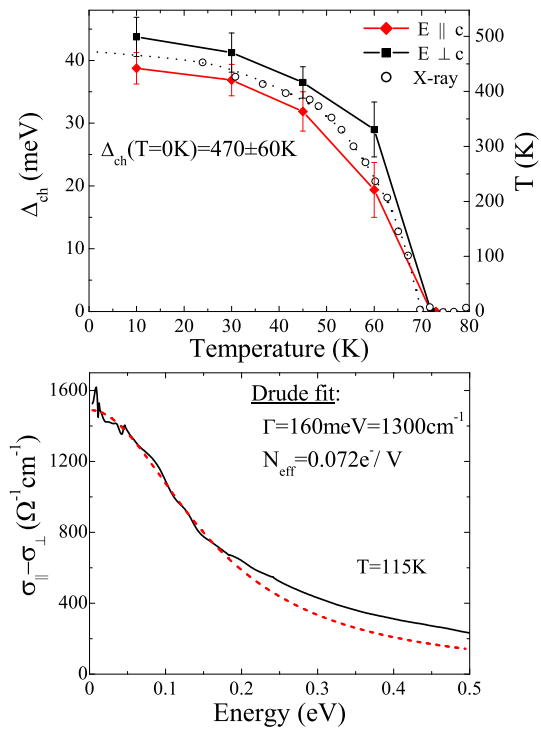}
\caption{(Color online) Top panel: Temperature dependence of the
charge gap in the insulating phase of BaVS$_3$ as determined from
the optical conductivity (full symbols), and the structural order
parameter (open symbols, on an arbitrary vertical scale) reproduced
from Ref.~\onlinecite{Inami,Fagot}. The dashed line corresponds to
the BCS gap profile. Bottom panel: Low-energy conductivity of the
$A_{1g}$-type V$(3d)$ electrons in the metallic phase of BaVS$_3$ at
$T=115$\,K evaluated according to
$\sigma_{A_{1g}}=\sigma_{\parallel}-\sigma_{\perp}$ (see text for
details). The dashed line indicates the Drude fit.}
\end{figure}

The separation of the different orbital contributions to the optical
conductivity indicates weak coupling between the $A_{1g}$ and $E_g$
electrons in the "bad-metal" phase of BaVS$_3$. This picture is
supported by the magnetic susceptibility which follows the
Curie-Weiss law above $T_{MI}$, hence it is fully attributed to the
localized $E_g$ electrons. Fig.~1 shows that there is no remarkable
deviation from the linearity in $1/\chi(T)$ down to the close
vicinity of the MIT, indicating the absence of precursor
fluctuations in the $E_g$ spin sector. In contrast, the in-chain
transport coefficients governed by the $A_{1g}$-type electrons fully
reflect the $1D$ structural fluctuations in the broad range
$T_{MI}<T<140$\,K. On this basis, we conclude that the primary force
of the transition is the $1D$ nature of the compound and thus the
formation of CDW in the $A_{1g}$ channel. The first manifestation of
a finite coupling between the two kinds of electrons is the
susceptibility cusp at $T_{MI}$ showing that $E_g$ electrons are
also involved in the transition. The present results clearly
evidence that the two sub-systems undergo a simultaneous transition,
even with a similar gap value in the ground state. The MIT is
accompanied by a large spectral weight transfer in
$\sigma_{A_{1g}}(\omega)$ from the Drude term to the correlation
peak (the sharp structure centered at $\sim0.25$\,eV corresponding
to excitations through the gap), while the small spectral weight
shifted by the opening of the gap in $\sigma_{E_g}(\omega)$ is
redistributed over a broader range of energy. The freezing out of
$1D$ lattice fluctuations, i.e. the lack of a subgap tail in
$\sigma_{\parallel}(\omega)$ indicates the role of $E_g$ electrons
in stabilizing the insulating state.

Recent X-ray experiments indicate that even in the tetramerized
phase the valence of the vanadium ions is uniform, i.e. V$^{4+}$
\cite{structure}. Apparently, the charge modulation in the $A_{1g}$
channel is compensated by $E_g$ electrons. In a sense, this is
natural in a system where the Coulomb correlation energies are large
($\sim1-2$\,eV), and even the localized electrons has an orbital
degree of freedom. Since the approximately two $E_g$ electrons in
the tetramerized unit cell preferably occupy neighboring sites
around the minima of the density wave, they will be susceptible of
singlet formation. This "confinement of $E_g$ electrons by the CDW"
explains the drop of the susceptibility below $T_{MI}$, and the
development of a spin gap of $\Delta_s\approx20$\,meV
\cite{Nakamura2000, Nakamura1997,Kezsmarki2001}. The weak coupling
between $E_g$ spins in the metallic state (reflected in the small
Weiss temperature $\Theta< 10$\,K$\ll T_{MI}$) is in agreement with
this picture: they have a tendency to form singlets, but the actual
phase transition is not governed by this energy scale.

In conclusion, we have studied the charge dynamics in the vicinity
of the correlation-driven MIT in the vanadium chain compound
BaVS$_3$ by polarized infrared spectroscopy. The unambiguous
separation of the different orbital contributions to the low-energy
optical conductivity in the metallic state allows the
characterization of two types of electrons coexisting around the
Fermi energy. The quasi $1D$ metallic transport of $A_{1g}$
electrons along the chains is superimposed on the isotropic hopping
conduction of localized $E_g$ electrons. The "bad-metal" character
and the weak anisotropy are the consequence of the large effective
mass ($m_{eff}\approx7m_e$) and the ill-defined nature of the
quasi-particles in the $A_{1g}$ conduction band. The latter is
clearly manifested in the anomalously high scattering rate
$\Gamma\geq160$\,meV almost comparable to the bandwidth. For
$T_{MI}<T<100$\,K the onset of a pseudo-gap is observed while in the
insulating phase the charge gap follows the temperature dependence
of the structural order parameter as the $1D$ lattice fluctuations
are quenched. We believe that a weakly coupled model of itinerant
$A_{1g}$ and localized $E_g$ electrons can give a comprehensive
description of BaVS$_3$.

We are indebted to M. Miljak for the susceptibility measurements.
Discussions with P. Fazekas, P. Thomas, I. Kup\^ci\'c and S.
Bari\^si\'c are gratefully acknowledged. L.M. received the
Szent-Gy\"orgyi Fellowship from the Hungarian Ministry of Education
and I.K. was a grantee of Bolyai J\'anos Fellowship. This work was
supported by the Hungarian Research Founds OTKA TS049881, by the
Swiss NSF, and its NCCR "MaNEP" and by the DOE under contract number
DE-AC02-98CH10886.
%

\end{document}